\documentclass[journal,twocolumn]{IEEEtran}

\ifCLASSINFOpdf
  \usepackage[pdftex]{graphicx}
  \graphicspath{{fig/}}
  \DeclareGraphicsExtensions{.pdf,.jpeg,.png}
\else
  \usepackage[dvips]{graphicx}
  \graphicspath{{fig/}}
\fi
\ifCLASSOPTIONcompsoc
  \usepackage[caption=false,font=normalsize,labelfont=sf,textfont=sf]{subfig}
\else
  \usepackage[caption=false,font=footnotesize]{subfig}
\fi

\usepackage[cmex10]{amsmath}
\usepackage{amsfonts}
\usepackage{amssymb}
\usepackage{algorithm}
\usepackage{algpseudocode}
\usepackage{subfig}
\usepackage{color,soul}
\usepackage{url}
\usepackage{epstopdf}
\usepackage{mwe}
\usepackage[export]{adjustbox}
\usepackage{booktabs}
\usepackage{xcolor,colortbl}
\usepackage{bm}
\usepackage{todonotes}
\usepackage{comment} % to comment out big sections or for backup of tex purposes
\usepackage{multirow}
\begin{document}

\title{Deep Learning Based Detection and Correction of Cardiac MR Motion Artefacts During Reconstruction for High-Quality Segmentation}

\author{ Ilkay~Oksuz, James~R.~Clough, Bram Ruijsink, Esther~Puyol-Ant\'on, Aurelien Bustin, Gastao Cruz, Claudia Prieto, Andrew P. King, Julia A. Schnabel

\thanks{Copyright (c) 2019 IEEE. Personal use of this material is permitted. However, permission to use this material for any other purposes must be obtained from the IEEE by sending a request to pubs-permissions@ieee.org. 

This work was supported by an EPSRC programme Grant (EP/P001009/1) and the Wellcome EPSRC Centre for Medical Engineering at the School of Biomedical Engineering and Imaging Sciences, King’s College London (WT 203148/Z/16/Z). Ilkay Oksuz was funded by the Scientific and Technological Research Council of Turkey (TUBITAK) grant number 118C353. This research has been conducted using the UK Biobank Resource under Application Number 17806. The GPU used in this research was generously donated by the NVIDIA Corporation. Correspondence to I. Oksuz.}     

\thanks{I. Oksuz is with Computer Engineering Department, Istanbul Technical University, Istanbul, Turkey}
\thanks{I. Oksuz, J. R. Clough, Bram Ruijsink, Esther~Puyol-Ant\'on, Aurelien Bustin, Gastao Cruz, Claudia Prieto, Andrew P. King, Julia A. Schnabel are with School of Biomedical Engineering and Imaging Sciences, King's College London, London, UK e-mail: (ilkay.oksuz@kcl.ac.uk).}

}
\maketitle

% ********* PAPER CORE *********

\begin{abstract}

\boldmath Segmenting anatomical structures in medical images has been successfully addressed with deep learning methods for a range of applications. However, this success is heavily dependent on the quality of the image that is being segmented. A commonly neglected point in the medical image analysis community is the vast amount of clinical images that have severe image artefacts due to organ motion, movement of the patient and/or image acquisition related issues. In this paper, we discuss the implications of image motion artefacts on cardiac MR segmentation and compare a variety of approaches for jointly correcting for artefacts and segmenting the cardiac cavity. The method is based on our recently developed joint artefact detection and reconstruction method, which reconstructs high quality MR images from k-space using a joint loss function and essentially converts the artefact correction task to an under-sampled image reconstruction task by enforcing a data consistency term. In this paper, we propose to use a segmentation network coupled with this in an end-to-end framework. Our training  optimises three different tasks:  1) image artefact detection, 2) artefact correction and 3) image segmentation.  We train the reconstruction network to automatically correct for motion-related artefacts using synthetically corrupted cardiac MR k-space data and uncorrected reconstructed images. Using a test set of 500  2D+time cine MR acquisitions from the UK Biobank data set, we achieve demonstrably good image quality and high segmentation accuracy in the presence of synthetic motion artefacts. We showcase better performance compared to various image correction architectures.
\end{abstract}

\ifCLASSOPTIONpeerreview
\else
\fi
\IEEEpeerreviewmaketitle

\section{Introduction}
\label{sec:introduction}

\IEEEPARstart{I}{mage} reconstruction, image denoising and downstream tasks (e.g. registration and segmentation) are traditionally considered as three separate tasks in medical image analysis. Medical image analysis techniques are typically applied to the raw data in a serial fashion, which can introduce a cascade of errors from one task to the next, particularly when the quality of the acquired data is low. 
While medical image analysis is taking an increasingly important role in clinical decision making, an often neglected step in automated image analysis pipelines is the assurance of image quality.
This is an important step because high accuracy in downstream tasks such as segmentation depends strongly on high quality medical images \cite{Oksuz2019b}. 

% CMR
Cine cardiac magnetic resonance (CMR) images are routinely acquired for the assessment of cardiac health, and can be used to derive metrics of cardiac function including volume, ejection fraction and strain \cite{Ruijsink2019}, as well as to investigate local myocardial wall motion abnormalities. 
CMR images are often acquired in patients who already have existing cardiovascular disease. These patients are more likely to have arrythmias or have difficulties with either breath-holding or remaining still during acquisition. Therefore, the images can contain a range of image artefacts \cite{Ferreira2013}, and assessing the quality of images acquired by MR scanners is a challenging problem.  
Misleading segmentations can be the result and can lead clinicians to draw incorrect conclusions from the imaging data \cite{Oksuz2019c}.
In current clinical practice, images are visually inspected by one or more experts, and those of insufficient quality are excluded from further analysis and/or reacquired. However, if successful image correction and segmentation algorithms are in place, these data could be utilised for further clinical evaluation.

% Overview of paper
In this paper, we propose a deep learning based approach for a fully automated framework for joint motion artefact detection, correction and segmentation in cine CMR short axis images. 
A novel end-to-end training setup is proposed to detect and correct motion artefacts and extract segmentations for the corrected images in a comprehensive, integrated framework. 
An analysis of multiple deep learning architectures and learning mechanisms is also presented. 
This paper builds upon our previously presented work \cite{Oksuz2019c}, in which we proposed the use of a Convolutional Neural Network (CNN) architecture to both detect and correct motion artefacts. 
Here, we extend this idea to a joint training approach for image artefact detection, correction and segmentation. 

\begin{figure}[htbp]
\begin{minipage}[b]{0.23\linewidth}
  \centering
  \centerline{\includegraphics[width=\linewidth]{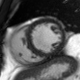}}
%  \vspace{1.5cm}
  \centerline{(a)  }\medskip
  \label{fig:Motivationa}
\end{minipage}
\hspace{0.01cm}
\begin{minipage}[b]{0.23\linewidth}
  \centering
  \centerline{\includegraphics[width=\linewidth]{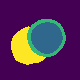}}
%  \vspace{1.5cm}
  \centerline{ (b) } \medskip
  \label{fig:Motivationc}
\end{minipage}
\hspace{0.01cm}
\begin{minipage}[b]{0.23\linewidth}
  \centering
  \centerline{\includegraphics[width=\linewidth]{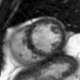}}
%  \vspace{1.5cm}
  \centerline{(c) }\medskip
  \label{fig:Motivationd}
\end{minipage}
\hspace{0.01cm}
\begin{minipage}[b]{0.23\linewidth}
  \centering
  \centerline{\includegraphics[width=\linewidth]{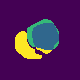}}
%  \vspace{1.5cm}
  \centerline{ (d) } \medskip
  \label{fig:Motivationf}
\end{minipage}

\caption{Examples of a (a) good quality cine CMR image; (b) segmentation output by a U-net \cite{Ronneberger2015}; (c) (synthetically) corrupted image; (d) segmentation output by the same network. Our figure illustrates the problems caused by motion artefacts for subsequent segmentation.}
\label{fig:Mot}
\end{figure}

Figure \ref{fig:Mot} illustrates the challenge image artefacts generate for state-of-the-art segmentation algorithms. The original image and the segmentation produced by a trained segmentation algorithm (U-net) are visualised in Figures \ref{fig:Mot}a and \ref{fig:Mot}b. A synthetically generated low quality image and its segmentation with the same network can be seen in Figures \ref{fig:Mot}c and \ref{fig:Mot}d. We aim to address  the issue of low quality segmentations from low quality image data, which is a result of problems in image acquisition.

The remainder of this paper is organised as follows. 
In Section \ref{sec:Related}, we first present an overview of the relevant literature in image artefact detection and correction.
Then, we review the literature on simultaneous image correction and downstream tasks, and present our novel contributions in this context. 
In Section \ref{sec:materials}, we provide details of the clinical data sets used. 
In Section \ref{sec:methods} we describe our proposed framework for image artefact detection, correction and segmentation including descriptions of the novel loss functions. 
Results are presented in Section \ref{sec:experiments_results}, while Section \ref{sec:discussion_conclusion} summarises the findings of this paper in the context of the literature and proposes potential directions for future work.

%%%%%%%%%%%%%%%%%%%%%%%%%%%%%%%%%%%%%%%%%%%%%%%%%%%%%%%%%%%%%%%%%%%%%%%%
%Related Works
%%%%%%%%%%%%%%%%%%%%%%%%%%%%%%%%%%%%%%%%%%%%%%%%%%%%%%%%%%%%%%%%%%%%%%%%

\section{Related works}
\label{sec:Related}

In this section, we provide an overview of the relevant literature on image quality assesment, image artefact correction and end-to-end training for downstream analysis of low quality images with a focus on applications in medical image analysis.

\subsection{Image quality assessment}
\label{sec:imq}

% Computer vision and Medical
Image quality assessment (IQA) has been an active area of research in computer vision and deep learning methods have shown great success on benchmark data sets \cite{Kang2014}. 
IQA is a vital step for analysing large medical image data sets as detailed in \cite{Chow2016}. 
Early efforts in medical imaging focused on quantifying the image quality of brain MR images \cite{Woodard2006}. The success of deep learning has motivated the medical image analysis community to utilise
such methods on multiple image quality assessment challenges such as fetal ultrasound \cite{Wu2017} and echocardiography %\citep{Abdi2017} 
\cite{Abdi2017} using 2D images and pre-trained networks. 
%Abdi et al. \cite{Abdi2017a} proposed to use temporal information using a Long Short Term Memory (LSTM) architecture to improve the accuracy of image quality assessment. 
Kuestner et al. \cite{Kuestner2018a} utilised a patch-based CNN architecture to detect motion artefacts in head and abdominal MR scans to achieve spatially-aware probability maps. 
In more recent work, \cite{Kuestner2018} proposed to utilise a variety of features and trained a deep neural network for artefact detection. \\
%They made use of an active learning strategy to detect low quality images due to the lack of sufficient low quality training data.\\

% CMR
CMR image quality issues have mostly been studied in the context of missing slices \cite{Zhang2017}, due to their adverse influence on calculating ventricular volume, and therefore, ejection fraction. 
Tarroni et al. \cite{Tarroni2018} proposed the use of a decision forest approach for heart coverage estimation, inter-slice motion detection and image contrast estimation in the cardiac region.  
CMR image quality has also been linked with automatic quality control of image segmentation in \cite{Robinson2019}. In a comprehensive segmentation framework, Alba et al. proposed to use a random forest classifier to eliminate failed myocardial segmentation \cite{Alba2018}. The authors of \cite{Lorch2017} investigated synthetic motion artefacts and used histogram, box, line and texture features to train a random forest algorithm to detect different artefact levels. 
In a more recent work, Oksuz et al. \cite{Oksuz2019a} proposed to use a curriculum learning strategy exploiting different levels of k-space corruption to detect cardiac motion artefacts. 
All of these techniques only detect the artefacts and do not correct the artefact-corrupted data. Therefore, they can be used as a rejection mechanism in clinical scenarios but cannot allow utilisation of corrupted data.

\subsection{Image artefact correction}
\label{sec:dataimbalance}
To make full use of available clinical databases, image artefact correction methods are instrumental as they help ensure that all images meet the same high standard of image quality. To the best of our knowledge, we are not aware of any method that is specifically developed for the task of correcting mis-triggering artefacts.
The use of deep learning based image reconstruction to increase the quality of MR imaging under accelerated acquisition has been of recent interest to the community \cite{Han2018}. 
Deep learning has recently shown great promise in the reconstruction  of highly undersampled MR acquisitions with CNNs \cite{Qin2018,Schlemper2017}. In a pioneering work, Schlemper et al. \cite{Schlemper2017} proposed to use a deep cascaded network to generate high quality images, and  Hauptmann et al.~\cite{Hauptmann2019}  proposed to use a residual U-net  to reduce aliasing artefacts due to undersampling, with the purpose of accelerating image acquisition.  Our work aims to exploit the advances made in these methods in the context of artefact-corrupted data, and to further extend the framework to encompass the common downstream task of image segmentation.

\subsection{End-to-end techniques}

The idea of addressing the problems of image quality and coupling them with downstream tasks (e.g. segmentation) has been proven to be effective in the computer vision literature \cite{Liu2017a}. Application-driven image denoising networks can even be trained without ground truth segmentations as illustrated in \cite{Wang2019}.

% Segmentation in presence of pathology and artefacts
Deep learning techniques have been utilised  for segmentation problems with high success \cite{Litjens2017}. However, the influence of variability in acquisition protocols, pathology and image artefacts is an often overlooked problem. In a recent work, Shao et al.~\cite{Shao2018} demonstrated the shortcomings of deep learning in the presence of pathology for brain MR segmentation with an emphasis on the selection of training data. Mortazi et al. illustrated the superior performance of multi-view CNNs on atrial segmentations \cite{Mortazi2017}.

In particular for medical imaging, combining reconstruction and segmentation has proven to be effective for undersampled k-space image acquisitions for CMR \cite{Huang2019} and brain MR \cite{Huang2018}. Schlemper et al. \cite{Schlemper2018} proposed a technique that can generate segmentations using fewer than 10 k-space lines. 
Similar techniques have been proposed to fuse super-resolution and CMR segmentation \cite{Oktay2018} by using anatomical constraints through auto-encoders. More recently, Sun et al. \cite{Sun2018} proposed using multiple cascaded blocks with the same encoder to simultaneously reconstruct and segment brain MR from under-sampled k-space acquisitions. Our work differs from these methods, which focus on undersampling for acceleration. We focus on artefacts and aim for improved quality of images and segmentations. 
%We want to leverage the redundant image information that is available in for tasks of image artefact detection, correction and segmentation. 
To the best of our knowledge, our work is the first that has investigated joint image artefact detection, correction and segmentation in a single pipeline for CMR.  We demonstrate application of the framework to CMR but we believe it could also offer benefits to brain MR or other applications.

\subsection{Contributions}
\label{sec:contributions}

There are four major contributions of this work:
\begin{itemize}
    \item To the authors' knowledge, this is the first paper that detects, corrects and segments CMR images with motion artefacts in a unified framework directly from k-space;
    \item We present an extensive analysis of segmentation of low quality CMR images with various deep learning architectures;
    \item  When used on k-space data that were originally of high quality our framework does not diminish the quality of reconstructed images or segmentations, and when used on artefact-corrupted k-space data it increases the quality of both images and segmentations;
    \item We prospectively analyse performance on a testing case to illustrate the potential of our technique for image artefact correction and segmentation.
\end{itemize}

This paper builds upon two previous works: 1) our unified network for image artefact detection and correction \cite{Oksuz2019c},  and 2) our investigation of the influence of low quality images on the final segmentation result using a comparison of various artefact correction strategies \cite{Oksuz2019b}. Here, we extend the idea of detecting and correcting the image artefacts in a single framework by incorporating a segmentation network and performing joint training of our pipeline.

%%%%%%%%%%%%%%%%%%%%%%%%%%%%%%%%%%%%%%%%%%%%%%%%%%%%%%%%%%%%%%%%%%%%%%%%
% Materials
%%%%%%%%%%%%%%%%%%%%%%%%%%%%%%%%%%%%%%%%%%%%%%%%%%%%%%%%%%%%%%%%%%%%%%%%
\section{Materials}
\label{sec:materials}

In this section, we detail the data set that is used to train and test our framework. We also discuss the synthetic image artefact generation mechanism to establish coupled low quality and high quality data to train our method.

We evaluate our approach using a subset of the UK Biobank data set \cite{Petersen2016}.
The subset consists  of  short-axis cine CMR images of 4000 subjects and was chosen so as to exclude any images with quality issues such as image artefacts or missing axial slices and was visually verified by an expert cardiologist. The short-axis images have an in-plane image resolution of $1.8 \times 1.8 $mm$^{2}$  with a slice thickness of 8.0 mm and a slice gap of 2 mm. A short-axis image stack typically consists of approximately  10-12 image slices and covers the full heart, and we selected the mid-ventricular slice for use in our experiments. In the UK Biobank data set, each cardiac cycle contains images at 50 time frames, but in this paper we used every other frame (25 frames in total) for analysis. 
All images were cropped to $176 \times 132$  image size, centred at the myocardium using the technique described in \cite{Oksuz2019a} to have consistent image matrices.  Details of the image acquisition protocol can be found in \cite{Petersen2016}. \\

\subsection{Synthetic phase generation}

Phase information is an important source of data in MR image reconstruction. However, the UK BioBank dataset contains only magnitude images and the absence of phase would change the nature of simulated motion artefacts (e.g., decoherence from motion artefacts with dynamic phase occurs in practice but would not appear in the simulated artefact images). Therefore, we utilised a similar strategy to \cite{Haldar2013} in order to generate synthetic phase. Briefly, we first add white Gaussian noise to the original images, then apply a Fourier Transform to generate k-space data. The synthetic phase is generated by applying a low pass filter to the k-space data followed by an inverse Fourier transform.

\subsection{Synthetic image artefact generation}
\label{sect:artefactgen}

\begin{figure}[tb]
  \centering
  \centerline{\includegraphics[width=\linewidth]{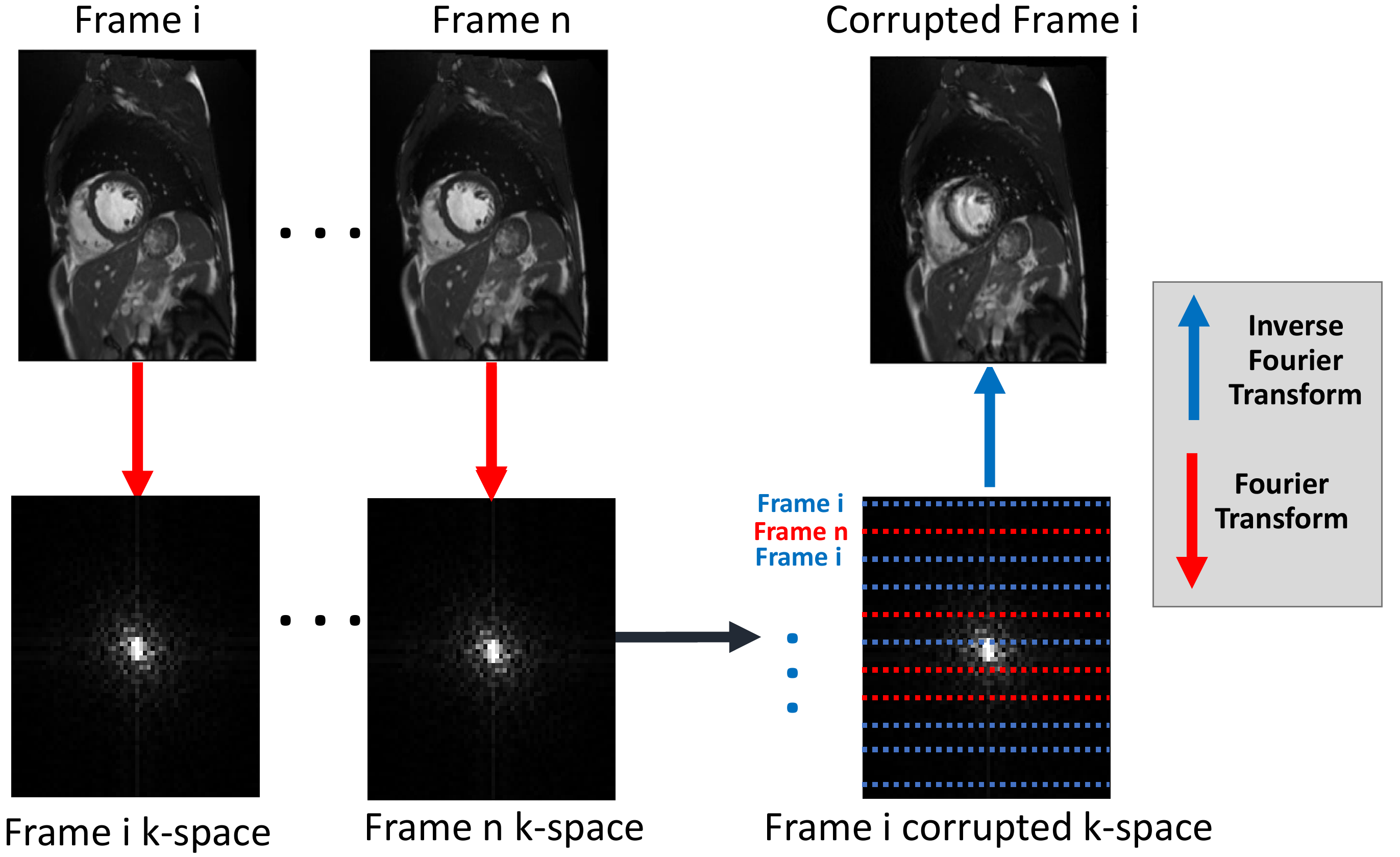}}
\caption{K-space corruption for synthetic motion artefact generation in k-space. The Fourier transform of each image frame is applied to generate the k-space representation of each image. We replace k-space lines with lines from different temporal frames to generate corruptions. }
\label{fig:Kspacemistrigg}
\end{figure}

From the high quality subset of the UK Biobank data set, we generated k-space corrupted data in order to simulate motion artefacts.
The UK Biobank data set was acquired using Cartesian sampling and we follow a Cartesian k-space corruption strategy to generate synthetic but realistic motion artefacts \cite{Oksuz2018b}. We first transform each 2D short axis sequence to the Fourier domain and change 1 in $z$ Cartesian sampling lines to the corresponding lines from other cardiac phases in order to mimic cardiac mistriggering artefacts, similar to \cite{Lorch2017}.  By using different values for $z={2,4,8,16,32}$, we are able to generate cardiac motion artefacts with different severities. In Figure \ref{fig:Kspacemistrigg} we show an example of the generation of a corrupted frame $i$ from the original frame $i$ using information from the k-space data of other temporal frames. We add a random frame offset $j$ when replacing the lines. This parameter is chosen from any of the other frames, drawn from a Gaussian distribution centered at the frame to be corrupted.\\

%%%%%%%%%%%%%%%%%%%%%%%%%%%%%%%%%%%%%%%%%%%%%%%%%%%%%%%%%%%%%%%%%%%%%%%%
% Methods
%%%%%%%%%%%%%%%%%%%%%%%%%%%%%%%%%%%%%%%%%%%%%%%%%%%%%%%%%%%%%%%%%%%%%%%%
\section{Methods}
\label{sec:methods}

In this section we first describe the unified framework for image artefact detection, correction and segmentation. Then, we provide details of the neural network architectures used for each task.  Finally, we introduce our joint loss function and the optimisation scheme.

\subsection{Architecture}
\label{sec:arc}

\begin{figure*}[htb]
  \centering
  \centerline{\includegraphics[width=\linewidth]{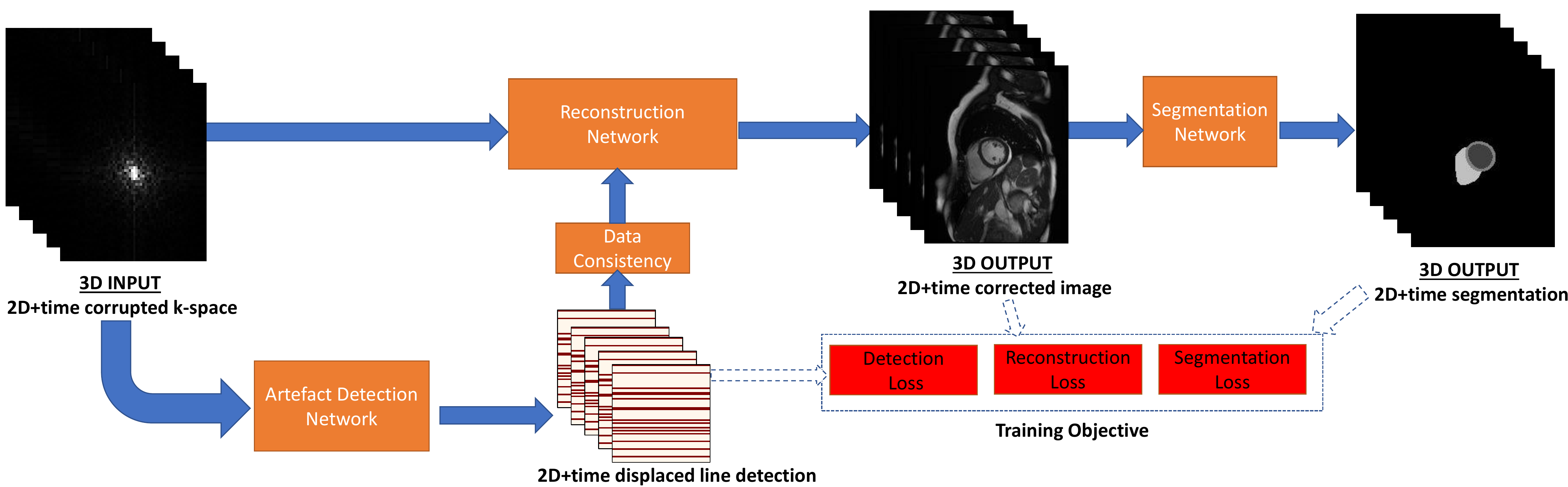}}
\caption{The 2D+time CNN architecture for motion artefact detection, reconstruction and segmentation. Our network detects and corrects image artefacts and outputs a motion corrected image sequence, which is also segmented by a segmentation network. The segmentation is back-propagated through the network, which further improves on the reconstruction in our training setup. }
\label{fig:Model}
\end{figure*}

The overall architecture of our network is illustrated in Figure \ref{fig:Model}. The proposed framework consists of 3 sub-networks and 3 corresponding loss function terms, which we detail below.

\subsection{Image motion artefact detection}
\label{sec:modelcorrection}

\begin{figure}[htb]
  \centering
  \centerline{\includegraphics[width=\linewidth]{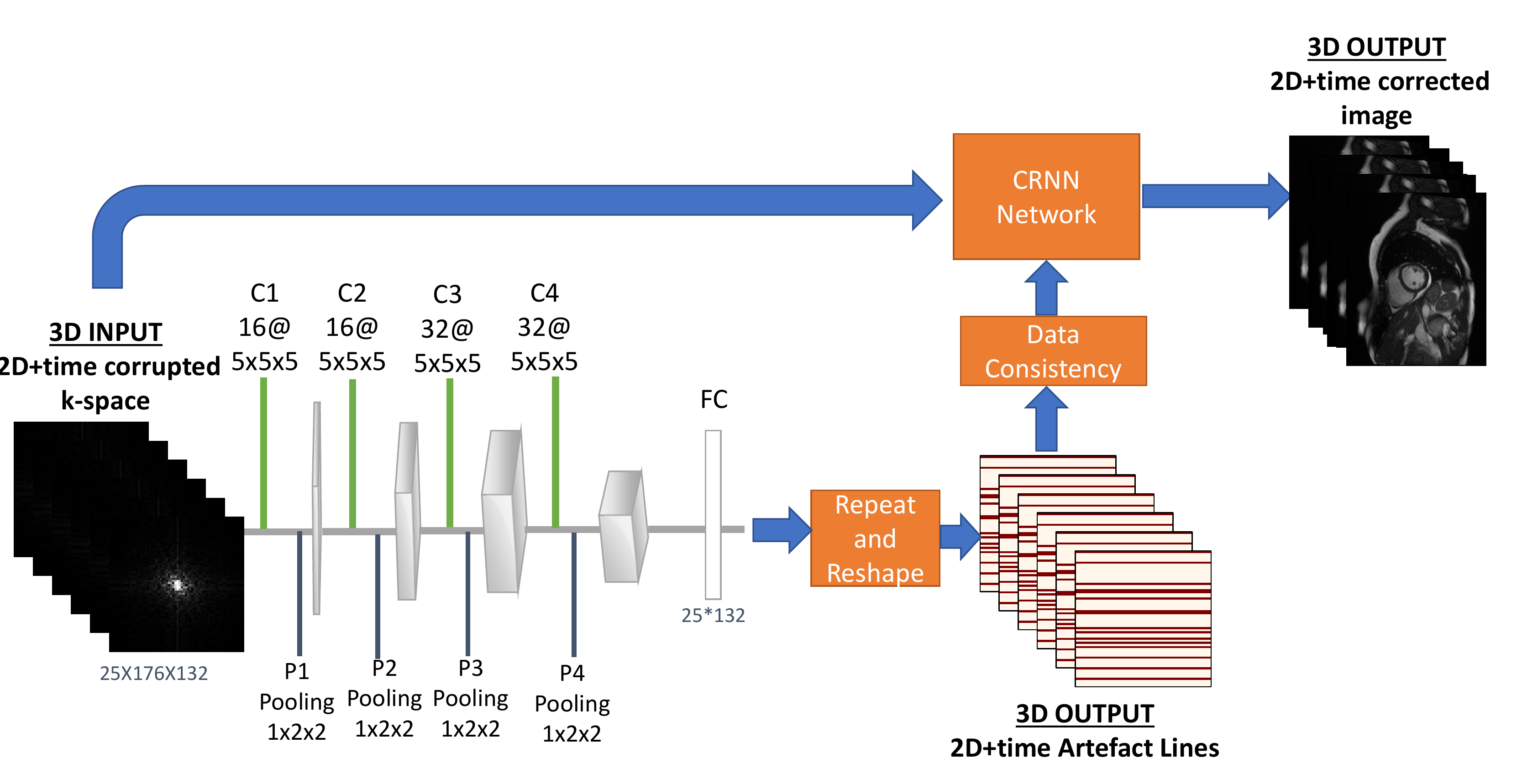}}
\caption{ The CNN architecture for motion artefact correction. Our network is able to detect displaced acquisitions in k-space and uses a hard data consistency term to  achieve high image quality.}
\label{fig:CorModel}
\end{figure}

The proposed artefact correction network architecture consists of two building blocks as visualised in Fig. \ref{fig:CorModel} and described in \cite{Oksuz2019c}: 1) a k-space line detection network to define the data consistency term; 2) a recurrent neural network architecture to correct image artefacts. All image data consist of complex numbers, which are treated as two channels in the network.

The proposed artefact detection CNN consists of eight layers as visualised in Fig. \ref{fig:CorModel}. The architecture of our network follows a similar architecture to the one in \cite{Tran2015}, which was originally developed for video classification using a spatio-temporal 3D CNN. In our case, we use the third dimension as the time component and 2D+time mid-ventricular sequences for classification of corrupted k-space lines. The input data are intensity normalised $176 \times 132$  CMR images with 25 time frames with complex input treated as an additional channel. The network has 4 convolutional layers and 4 pooling layers, 1 fully-connected layer and a softmax loss layer to predict artefact-corrupted k-space lines.  After each convolutional layer, a ReLU activation is used. We then apply pooling on each feature map to reduce the filter responses to a lower dimension. We  apply dropout with a probability of 0.2 at all convolutional layers and after the first fully connected layer to enforce regularisation. All of these convolutional layers are applied with appropriate padding of 2 and stride of 1. The final output of this network is a 1-dimensional vector with length equal to the number of Cartesian lines present in the k-space. Finally, we repeat these values and reshape them to a rectangular matrix with the same size as our 2D+time input to be able to use them in the hard data consistency term as a thresholded binary mask.

\subsection{Motion corrected image reconstruction }

In our algorithmic setup we utilise a convolutional recurrent neural network (CRNN) architecture \cite{Qin2018} as the reconstruction network. The CRNN is designed to reconstruct CMR images from undersampled k-space data by jointly exploiting the dependencies of the temporal sequences as well as the iterative nature of traditional regularised MR reconstruction. In addition, spatio-temporal dependencies are simultaneously learned by exploiting bidirectional recurrent hidden connections across time sequences. The network consists of a bi-directional recurrent neural network for exploiting temporal information, and convolutional recurrent layers to propagate information between iterations.

However, we note that our framework is flexible and, in principle, other published reconstruction networks could be used in place of this network. The CRNN network was chosen because of its capability to incorporate information from different temporal frames, which is instrumental in correcting k-space based artefacts. 10 iterations of the network were utilised as suggested in \cite{Qin2018}.

\subsection{Image Segmentation}
\label{sec:modelsegmnetation}

\begin{figure}[htb]
  \centering
  \centerline{\includegraphics[width=\linewidth]{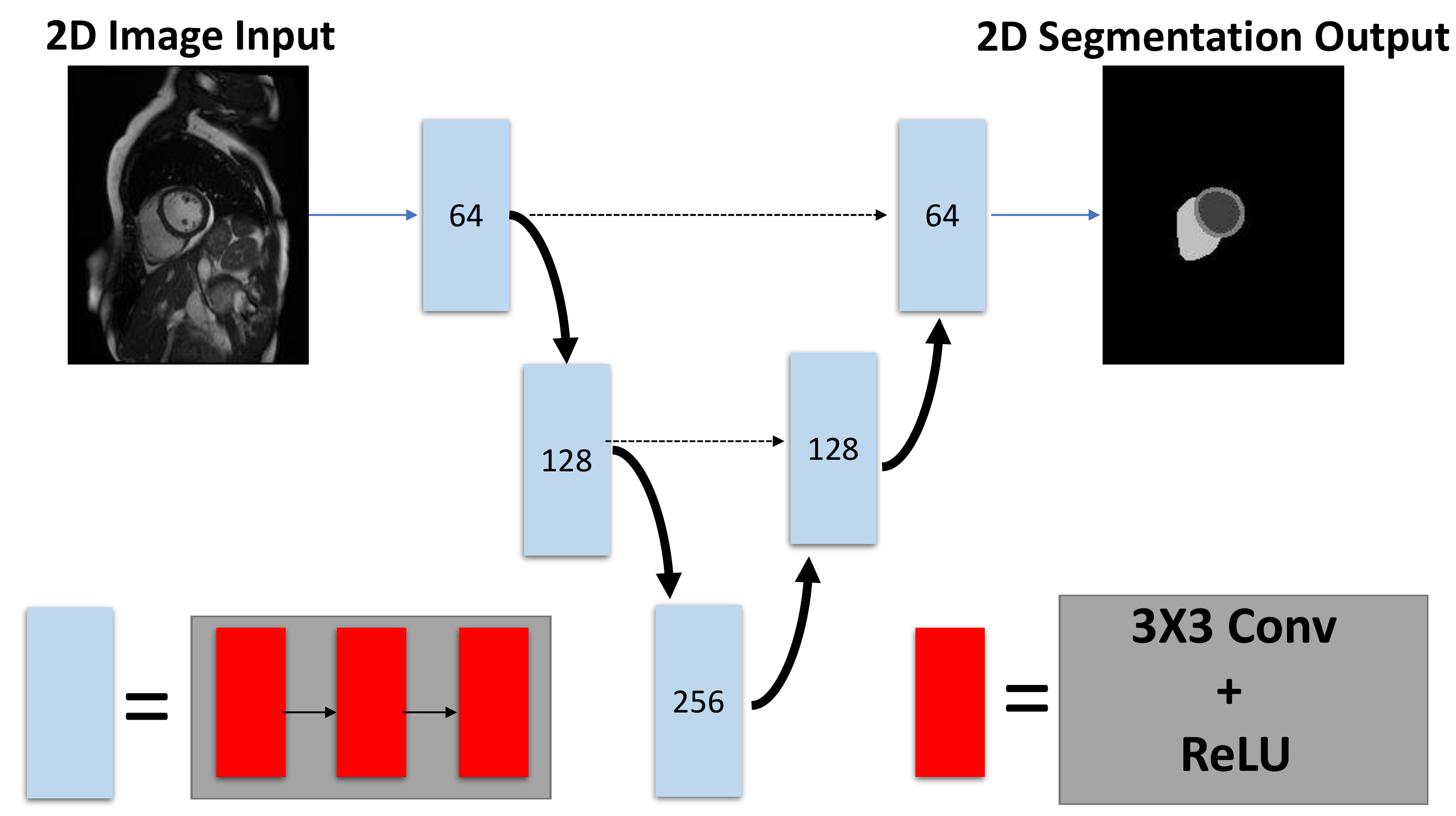}}
\caption{Diagram of the U-net architecture used for segmentation of 2D temporal sequences. Numbers correspond to the number of channels, and dotted arrows correspond to feature concatenation.}
\label{fig:UnetModel}
\end{figure}

Our segmentation network is a classical U-net segmentation network \cite{Ronneberger2015}, which is known to perform well on the mid-ventricular segmentation task in high quality images \cite{Bai2018}. The details of the network are illustrated in Fig. \ref{fig:UnetModel}. We chose to use a simple and well-established segmentation model in order to allow us to evaluate the influence of different strategies for combining motion-corrected reconstruction with segmentation (see Section \ref{sec:ArchitectureResults}).

% IO: Thanks JC for the figure idea

\subsection{Loss function}
\label{sec:loss}

Our loss function incorporates terms from all three sub-networks with the overall loss function defined as:

% Loss functions
$$\mathcal{L}_{\text{total}}=   (1-\lambda) \mathcal{L}_{\text{segmentation}} + \lambda   \mathcal{L}_{\text{correction}} $$

where  $ \mathcal{L}_{\text{correction}}$ refers to the joint image correction loss and $ \mathcal{L}_{\text{segmentation}}$ refers to the segmentation loss. $\lambda$ is a weighting parameter, which controls the influence of both terms on the total loss.

Our correction loss is a combination of image reconstruction loss and a cross-entropy loss:

$$\mathcal{L}_{\text{correction}}=   \gamma \mathcal{L}_{\text{detection}} +  (1-\gamma) \mathcal{L}_{\text{reconstruction}} $$

where $\gamma=0.3$, which is the optimised value for correction \cite{Oksuz2019c}.
The reconstruction loss is computed using the mean square error, defined as: 
$$\mathcal{L}_{\text{reconstruction}}=  \dfrac{1}{N_{p}} \sum_{p=1}^{N_{p}}  (I_{x}(p)-I_{y}(p))^2   $$

where  $N_{p}$ denotes the total number of pixels in images $x$ and $y$.
The detection loss is the cross entropy loss, defined as: 
$$\mathcal{L}_{\text{detection}}(pr,k)=  \dfrac{-1}{N_{l}} \sum_{l=1}^{N_{l}} (k \log(pr) + (1-k) \log(1-pr))$$
where $k$ is a binary indicator (0 or 1) indicating if a k-space line is corrupted or not and $pr$ is the predicted probability of the line being uncorrupted.  $N_{l}$ denotes the total number of k-space lines in an image.

Finally, the segmentation loss is the pixel-wise cross entropy loss:

$$\mathcal{L}_{\text{segmentation}}(y_{\text{pred}},y_{\text{true}})=  - \sum_{\text{classes}} y_{\text{true}} \log(y_{\text{pred}}) $$
where $y_{\text{pred}}$ is the probability of a segmentation label belonging to a class and $y_{\text{true}}$ is the ground truth segmentation label.

\subsection{Implementation details}
\label{sec:implementation}

During training, a batch-size of 10 2D+time sequences was used due to memory constraints. We used the Adam optimiser, whose momentum was set to 0.9 and the learning rate was  $5 \times 10^{-4}$.

The parameters of the convolutional and fully-connected layers were initialised from a zero mean, unit standard deviation Gaussian distribution. In each trial, training was continued until the network converged. Convergence was defined as a state in which no substantial progress was observed in the training loss. One percent improvement is considered significant in our training setup. Parameters were optimised using a grid-search among all parameters. 
We used the Pytorch framework for implementation. Training the network took around 15 hours on a NVIDIA Quadro P6000 GPU. Classification of a single 2D+time image  sequence took less than 1s.

%%%%%%%%%%%%%%%%%%%%%%%%%%%%%%%%%%%%%%%%%%%%%%%%%%%%%%%%%%%%%%%%%%%%%%%%
% Experiments and Results
%%%%%%%%%%%%%%%%%%%%%%%%%%%%%%%%%%%%%%%%%%%%%%%%%%%%%%%%%%%%%%%%%%%%%%%%
\section{Experiments and results}
\label{sec:experiments_results}

Four sets of experiments were performed. The first set of experiments (Section \ref{sec:ArchitectureResults}) aimed to compare the performance of different algorithmic approaches for automatic motion artefact correction and segmentation, while the second set of experiments (Section \ref{sec:CorrectionResults}) aimed at comparing different design choices for the correction network. To show the potential of our method as a global image reconstructor, we also report the results of each network using uncorrupted k-space data as input in Section \ref{sec:uncorrupted_results}. Finally, the experiments in Section \ref{sec:ProspectiveResults} validate the proposed network architecture for a prospective case study in which we utilise raw k-space data from a CMR acquisition.
Before describing the experiments in detail, we first describe the evaluation measures used.\\

\subsection{Evaluation metrics and methods of comparison}
\label{sec:error_measures}
Image quality is evaluated with Mean Absolute Error (MAE), Peak Signal to Noise Ratio (PSNR) and Structural Similarity Index (SSIM). For evaluating the prospective data, where no ground truth is available, the Sharpness Index (SI) \cite{Blanchet2012} is used to evaluate performance.

MAE is defined as:
$${\text{MAE}}=  \dfrac{1}{N_{p}} \sum_{p=1}^{N_{p}} | (I_{x}(p)-I_{y}(p)) |  $$
where p denotes each pixel and $N_{p}$ denotes the total number of pixels in images $I_{x}$ and $I_{y}$.

PSNR is defined as:
$${\text{PSNR}}=  20 \log_{10} (\text{max}(I)) - 10 log_{10}(\dfrac{1}{N_{p}} \sum_{p=1}^{N_{p}}  (I_{x}(p)-I_{y}(p))^2)    $$
where $ \text{max}(I)$ denotes the maximum intensity value in the ground truth image.

The SSIM between two images is  defined as follows for any image regions x and y:
$$\text{SSIM}(p)=  \dfrac{(2 \mu_{x}\mu_{y}+c_{1}) (2 \sigma_{xy}+c_{2})}{ (\mu_{x}^{2}+\mu_{y}^{2}+c_{1}) (\sigma_{x}^{2}+\sigma_{y}^{2}+c_{2})}  $$ 
where $\mu_{x}$ and $\mu_{y}$ are the average intensities for regions $x$ and $y$, $\sigma_{x}$ and $\sigma_{y}$ are variance values for regions $x$ and $y$, $\sigma_{xy}$ is the covariance of regions $x$ and $y$ and $c_{1}$ and $c_{2}$ are constant values for stabilising the denominator. 

The SI between two images is  defined as:

$$ SI(u) = − \log_{10} \Phi \frac{\mu - TV(u) }{\sigma}  $$

where $\mu  = E[TV (I)]$ is the expectation of the total variation of the image and $\sigma = Var [TV (I)]$ is the corresponding variance. 

We  computed the Dice overlap measure for evaluating segmentations, which is defined between two regions $A$ and $B$ as:

$$ D(A,B)=\dfrac{2 \|A \cap B \|}{\|A\| \cup \|B\|}. $$

\subsection{Architecture comparison}
\label{sec:ArchitectureResults}

\begin{figure}[htb]
  \centering
  \centerline{\includegraphics[width=\linewidth]{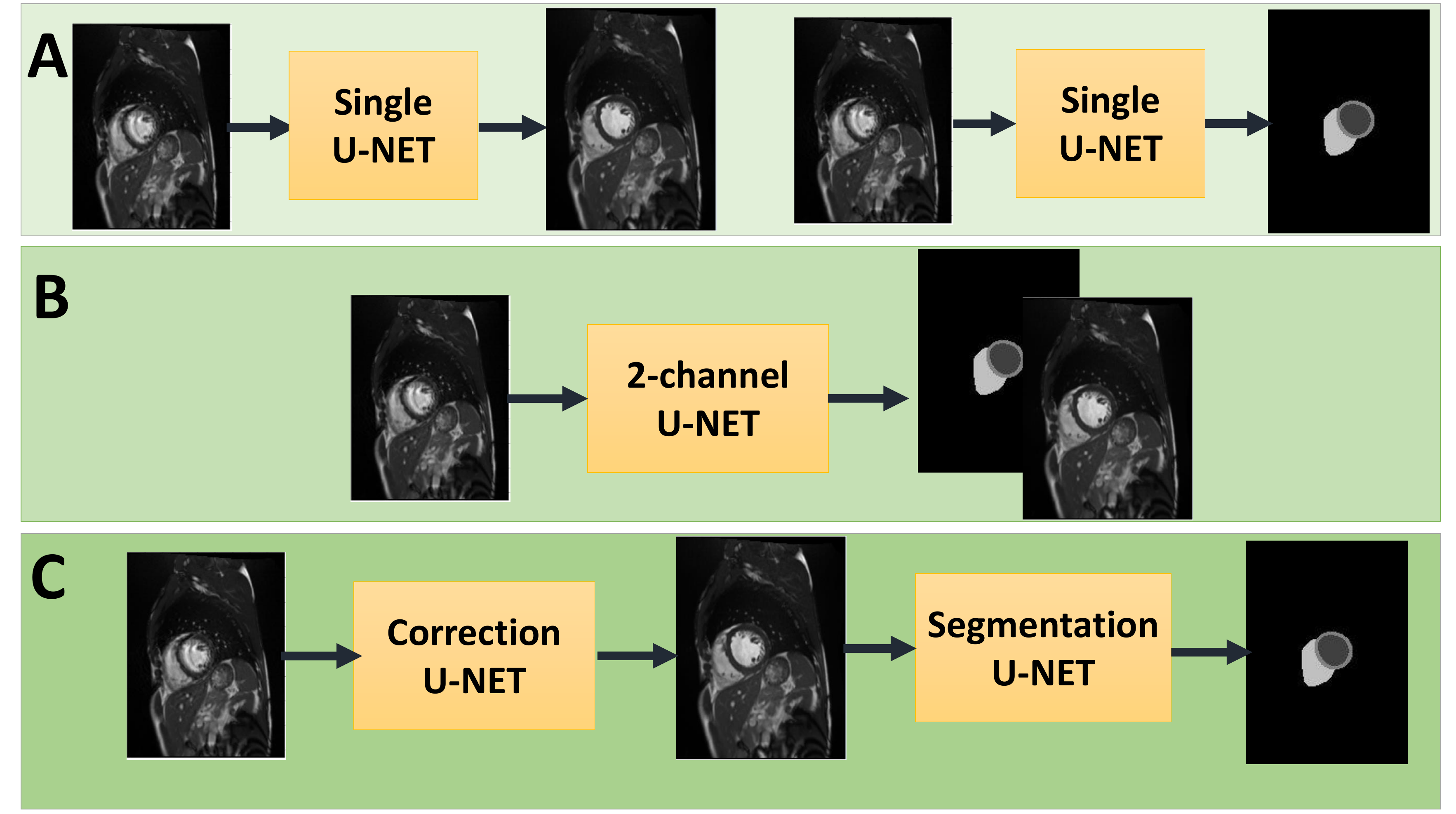}}
\caption{Compared architectures for image motion artefact correction and segmentation: (a) a single U-net, which addresses the tasks of correction and segmentation separately, (b) a U-net with 2-channel outputs, (c) two-consecutive networks, where first U-net corrects the image artefacts and the second one segments the corrected images. All architectures have low quality images as input.}
\label{fig:arccom}
\end{figure}

Two sets of tests were performed for deciding on the best architectural design to correct and segment CMR images with image artefacts. To decide on the best architecture, we tested several designs (as illustrated in Figure \ref{fig:arccom}), while keeping the total number of parameters in the models the same: 1) a single U-net for image correction and image segmentation (U-net); 2) a U-net with 2 output channels for the corrected image and the final segmentation output (U-net 2-channel); and 3) two serial U-nets to first output a corrected image and then segment the image (Cascaded U-nets). %\textcolor{blue}{Each model might have a different architecture, which in turn can generate different numbers of parameters and an unfair comparison. To adjust for the number of parameters we have changed the number of filters.}
To allow for a fair comparison, we adjusted the numbers of filters in each architecture to ensure that the numbers of model parameters were approximately equal.
 To evaluate all designs, we used a training set of 3000, a validation set of 500 and a test set of 500 2D+time CMR images from the UK Biobank subset described in Section \ref{sec:materials}.

 % Figure Image Quality (difference) and segmentation

% Architecture Segmentation Results Comment
In Table \ref{table:archseg}, we report the segmentation results of this experiment. 
We also report the best achievable result, which employs a single U-net model but does not use any corrupted images during training or testing (U-net-Top). Also, using the same model without further training we segmented the poor quality images to establish a baseline performance (U-net-Baseline). Our experiments show that the best segmentation can be achieved by using two serial networks and training them end-to-end, which motivates our design choice for the remaining experiments.

% Architecture Segmentation Table
\begin{table} 
\centering
\caption{Dice segmentation overlap of the algorithm outputs with the ground truth segmentation (manual delineation of the good quality image). The proposed architecture achieves high Dice overlap for the Left Ventricle  (LV), Myocardium (Myo) and Right Ventricle (RV) regions.}
\begin{tabular}{lcccc}
\hline
Methods              & LV    & Myo   & RV    \\
\hline
U-net-Top            & 0.985 & 0.958 & 0.944 \\
U-net-Baseline       & 0.932 & 0.865 & 0.873 \\
Single U-net         & 0.948 & 0.913 & 0.917 \\
U-net 2-channel      & 0.908 & 0.911 & 0.903 \\
\textbf{Cascaded U-nets}  & \textbf{0.962} & \textbf{0.918} & \textbf{0.920} \\

\hline
\end{tabular}
\label{table:archseg}
\end{table}

% Architecture Image Quality Results Comment
In Table \ref{table:archimg}, we report the image quality results of this experiment. In this table, we also report the original quality of the corrupted images as a baseline (Corrupted-Baseline). The results show that the best image quality can be achieved by using two serial networks and training them end-to-end, confirming the segmentation findings in Table \ref{table:archseg}. 

% Architecture Image Quality Results
\begin{table} 
\centering
\caption{Image Quality of the algorithm outputs with the good quality acquisition. The consecutive network architecture achieves the best image quality.}
\begin{tabular}{lccccc}
\hline
Methods              & MAE    & PSNR   & SSIM     & SI\\
\hline
Corrupted-Baseline & 0.068 & 19.156 & 0.751 & 55.292   \\ 
Single U-net       & 0.073 & 24.238 & 0.743 & 64.294   \\ 
U-net 2-channel    & 0.061 & 25.098 & 0.764 & 70.902  \\ 
\textbf{Cascaded U-nets} & \textbf{0.058} & \textbf{26.212} & \textbf{0.782} & \textbf{70.945}   \\ 

\hline
\end{tabular}
\label{table:archimg}
\end{table}

\subsection{Correction technique analysis}
\label{sec:CorrectionResults}
 
 % Correction Methods Tables comment
Having decided on the architecture to utilise for jointly addressing the reconstruction and segmentation tasks, we next aim to illustrate the benefit of our proposed technique. In Tables \ref{table:corseg} and \ref{table:corimg} we report the segmentation and image quality results respectively for this experiment. The results show that our novel technique, where we replace the correction U-net, improves image quality and final segmentation outputs.

% Correction Methods Segmentation Table
\begin{table} 
\centering
\caption{Dice segmentation overlap of the algorithm outputs with the ground truth segmentation (manual delineation of the good quality image). The proposed architecture achieves high Dice overlap for the Left Ventricle (LV), Myocardium (Myo) and Right Ventricle (RV) regions.}
\begin{tabular}{lcccc}
\hline
Methods                                  & LV    & Myo   & RV    \\
\hline
Cascaded U-nets                             & 0.962 & 0.918 & 0.920   \\
\textbf{Proposed}                    & \textbf{0.968}  &	\textbf{0.937}  &	\textbf{0.933}   \\            
\hline
\end{tabular}
\label{table:corseg}
\end{table}

% Correction Methods Image Quality Table
\begin{table} 
\centering
\caption{Image Quality of the algorithm outputs with the good quality acquisition. The proposed network can achieve high PSNR and SSIM.}
\begin{tabular}{lccccc}
\hline
Methods              & MAE    & PSNR   & SSIM   & SI\\
\hline
Cascaded U-nets              & 0.058& 26.212& 0.782  & 70.945   \\
\textbf{Proposed}     & \textbf{0.048}  &	\textbf{28.805}  &	\textbf{0.801}  & \textbf{75.819}   \\  
\hline
\end{tabular}
\label{table:corimg}
\end{table}

\begin{figure}[htb]
  \centering
  \centerline{\includegraphics[width=\linewidth]{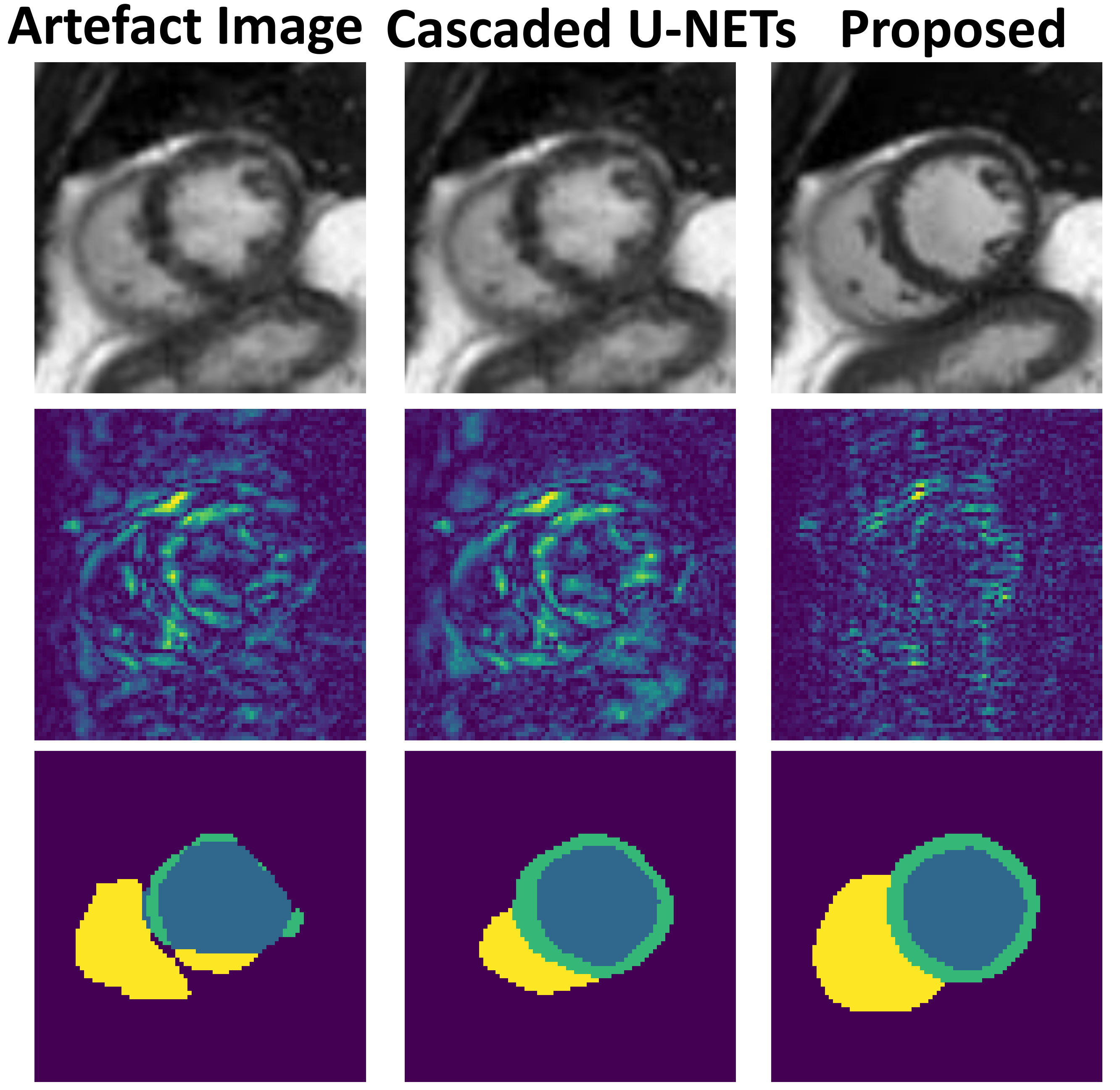}}
\caption{Example results for proposed architecture and cascaded U-nets. Top row shows a synthetic low quality cine CMR image (left), the corrected image output of 2 separately trained U-nets (middle) and the corrected image output of our proposed method (right). The second row shows the differences with corresponding good quality image for each case. The bottom row shows the generated segmentation outputs.} 
\label{fig:Results_difference}
\end{figure}

\begin{figure}[htb]
  \centering
  \centerline{\includegraphics[width=\linewidth]{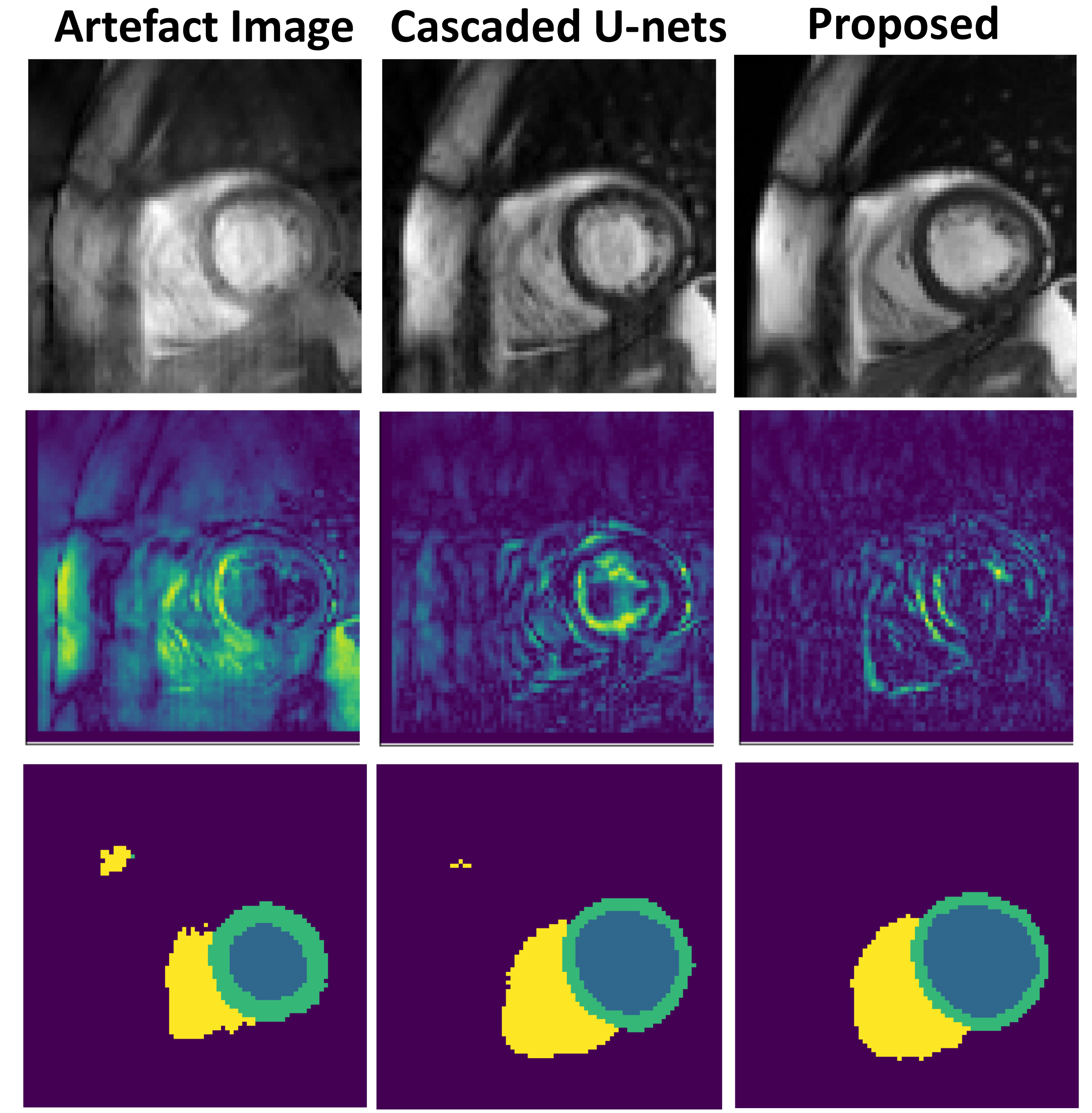}}
\caption{Example results for proposed architecture and cascaded U-nets. Top row shows a synthetic low quality cine CMR image (left), the corrected image output of 2 separately trained U-nets (middle) and the corrected image output of our proposed method (right). The second row shows the differences with corresponding good quality image for each case. The bottom row shows the generated segmentation outputs.} 
\label{fig:Results_difference2}
\end{figure}

 % Figure comment
%We evaluated different correction strategies in the serial training pipeline.
Results from this experiment are shown for two example cases in Figures \ref{fig:Results_difference} and \ref{fig:Results_difference2}.. The proposed technique can generate high quality images with low difference compared to the original images, with improvements in particular at edges. Moreover, the improved image quality results in higher quality segmentations, clearly delineating the myocardium, LV and RV with high accuracy. (More qualitative results are  provided in the supplementary material as videos.)

 % Figure Image Quality (difference) and segmentation

 % Figure Image Quality (difference) and segmentation

\subsection{Uncorrupted results}
\label{sec:uncorrupted_results}

We also used the trained models to test our framework on the original uncorrupted data, in order to demonstrate its potential as a general image reconstruction/segmentation tool. Table \ref{table:uncor} presents the results achieved with each algorithm. Our proposed technique diminishes the quality of images least and therefore shows great potential to be used as a global image reconstructor.

\begin{table*} [tb]
\centering
\caption{Mean image quality results of image quality correction for motion artefacts for uncorrupted inputs. Uncorrupted results use the correct k-space as input. The results indicate the potential of our method to be used as a global image reconstruction framework.}
\begin{tabular}{lccccccc}
\hline
& \multicolumn{3}{c}{Segmentation} & \multicolumn{4}{c}{Image Quality}  \\
\cline{2-4}
\cline{5-8}
Methods             & LV & Myo  & RV   & MAE & PSNR  & SSIM  & SI\\
\hline 

Single U-net            & 0.915 & 0.849 & 0.821  & 0.019 & 34.023 & 0.861 & 70.091   \\ 
U-net 2-channel         & 0.9358 & 0.861 & 0.857  & 0.017 & 35.922 & 0.863 & 73.182   \\ 
Cascaded U-nets         & 0.945 & 0.882 & 0.881 & 0.012 & 36.571 & 0.901 & 73.296   \\ 
\hline
\textbf{Proposed}   & \textbf{0.971} & \textbf{0.936} & \textbf{0.929}  & \textbf{0.004} & \textbf{39.17} & \textbf{0.949} & \textbf{84.193}    \\

\hline
\end{tabular}
\label{table:uncor}
\end{table*}

\subsection{Parameter analysis}

\begin{figure}[htb]
  \centering
  \centerline{\includegraphics[width=\linewidth]{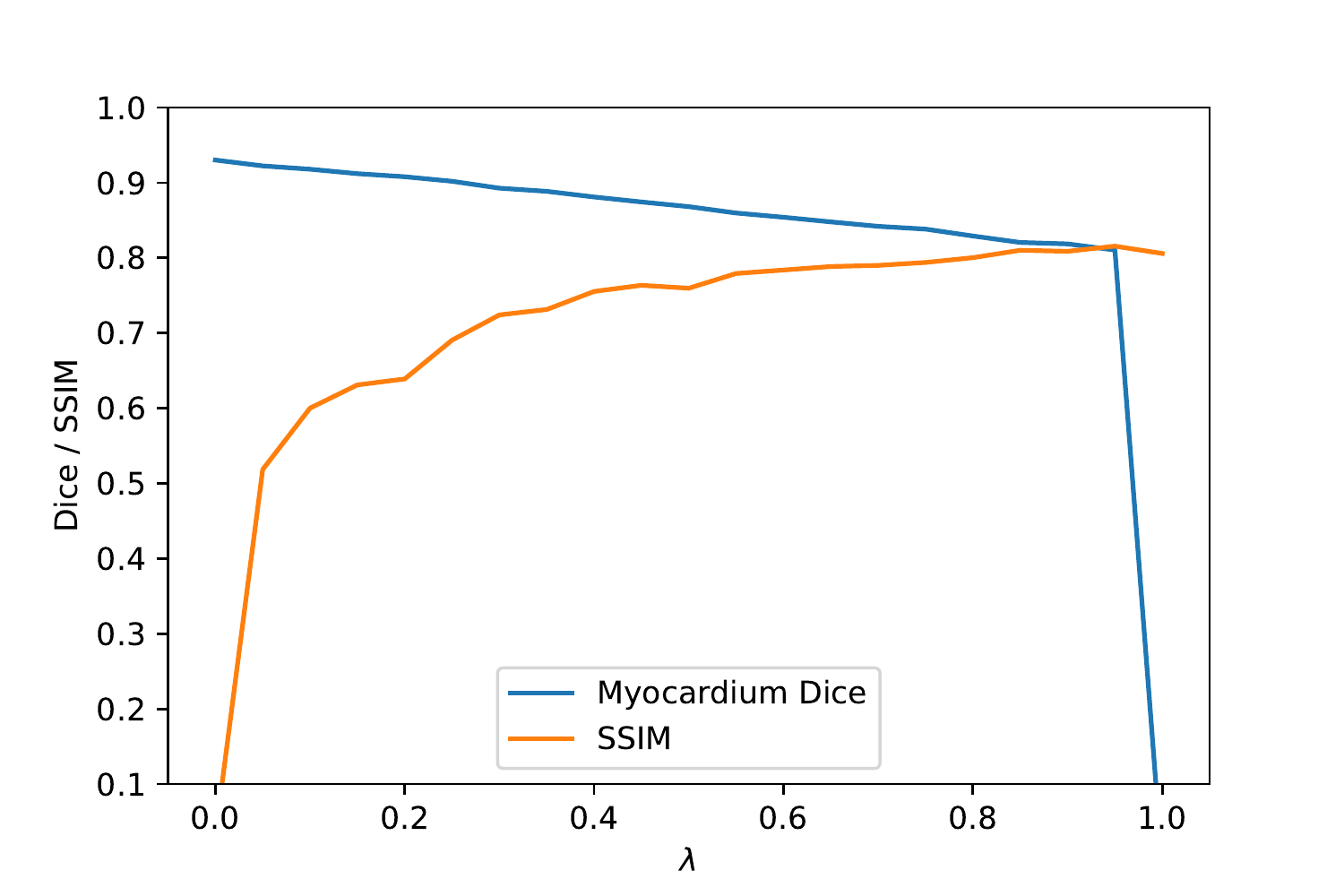}}
\caption{ Influence of the $\lambda$ parameter on segmentation accuracy (Dice) and image quality (SSIM). The weight plays an important role in adjusting our pipeline to put more emphasis on one of the two tasks.}
\label{fig:Lambda}
\end{figure}

Next, we repeated the training of our proposed technique for different $\lambda$ values to highlight the importance of this parameter on image quality and segmentation. We used a range of values from 0 to 1 in steps of 0.05, and the results are shown in Figure \ref{fig:Lambda}. We have used a $\lambda=0.8$ in our experiments. The parameter $\lambda$  weights the image reconstruction loss and segmentation loss and accordingly can be used to tune for either higher image quality or higher segmentation quality in our algorithmic framework depending on clinical choices.

\begin{figure}[htb]
  \centering
  \centerline{\includegraphics[width=\linewidth]{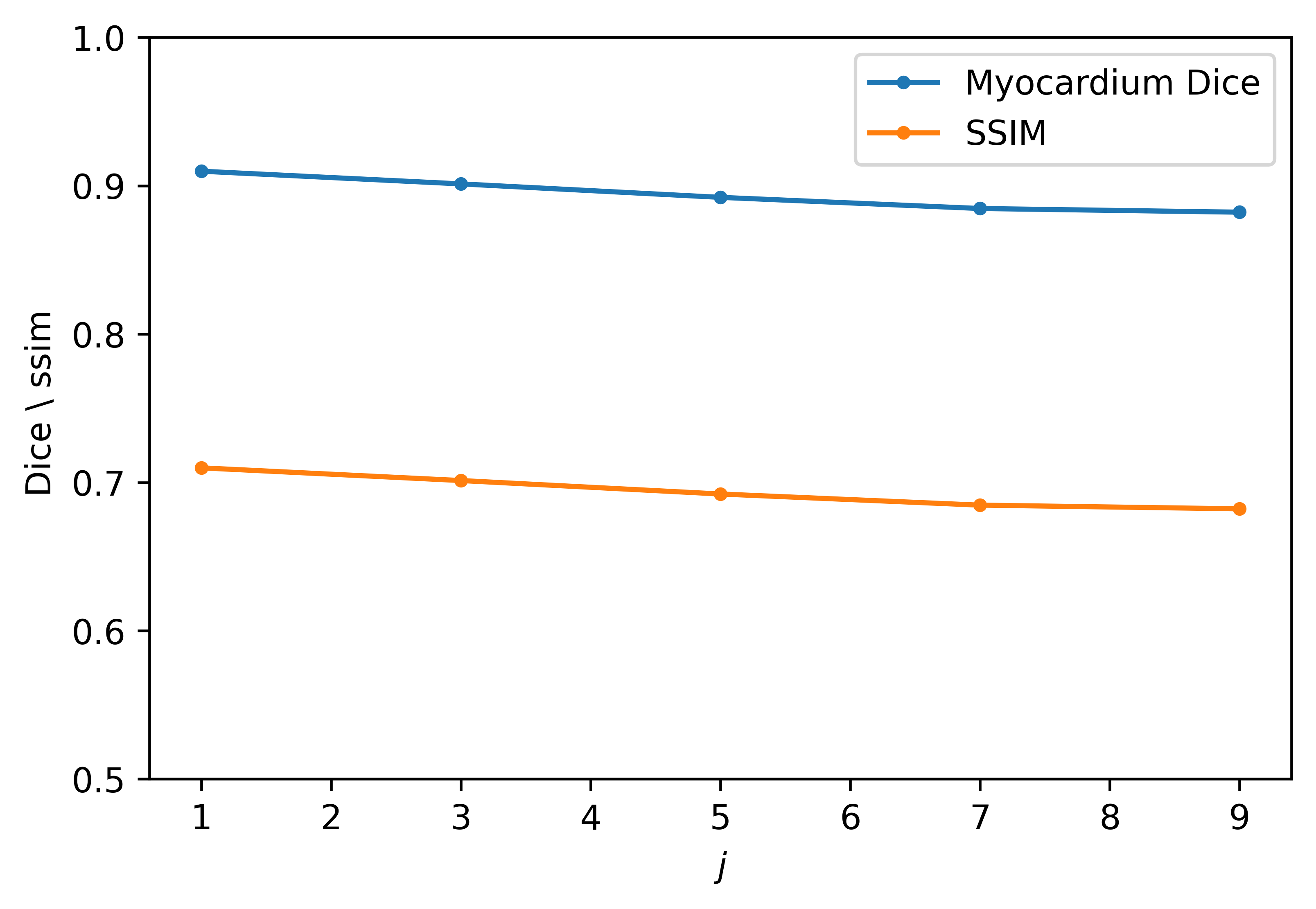}}
\caption{ Influence of the $j$ parameter (phase offset used to corrupt images) on segmentation accuracy (Dice) and image quality (SSIM). The increasing offset makes the task more challenging. The performance for random values are provided in Tables \ref{table:corimg} and \ref{table:corseg}. }
\label{fig:param_j}
\end{figure}

\begin{figure}[htb]
  \centering
  \centerline{\includegraphics[width=\linewidth]{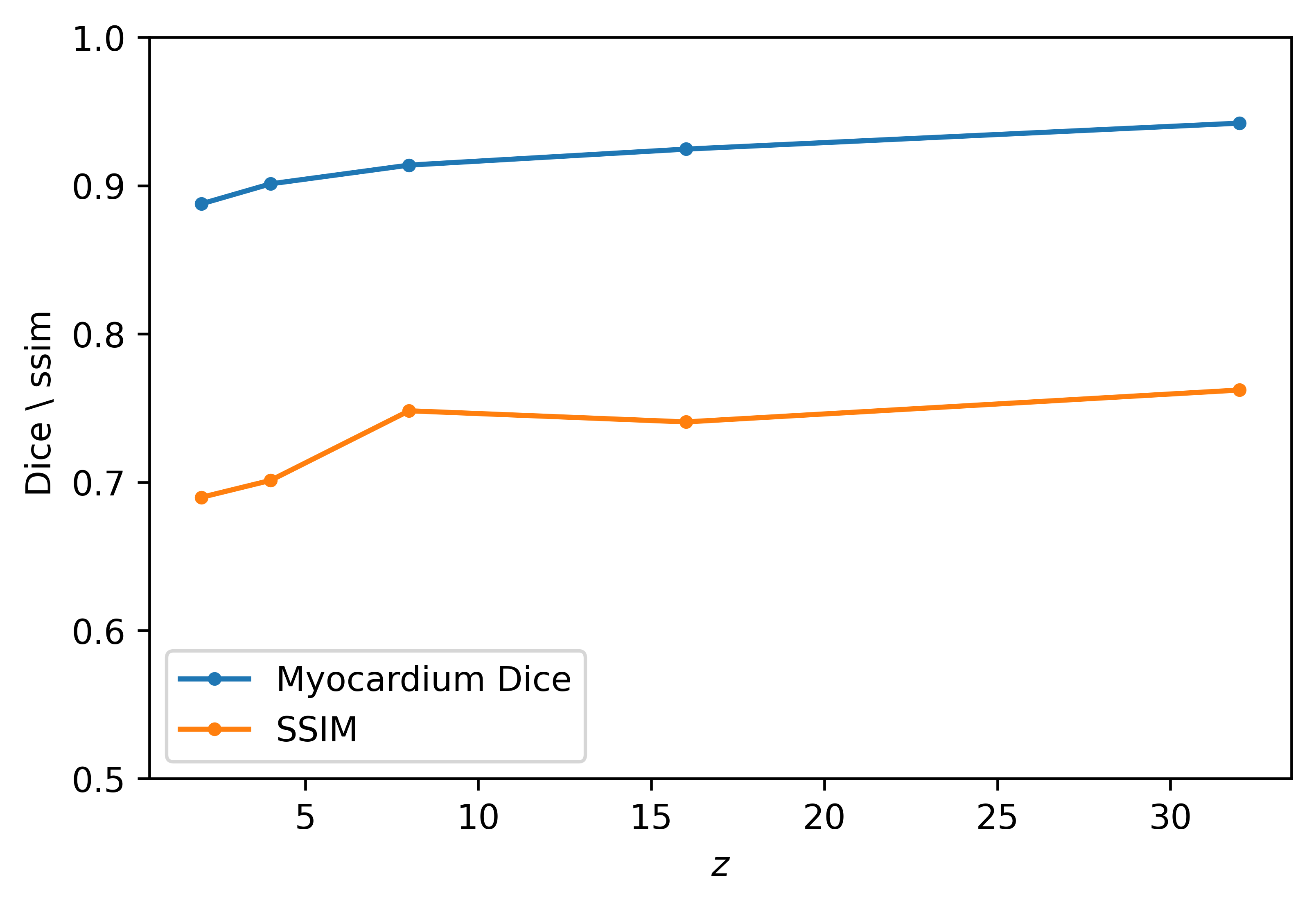}}
\caption{ Influence of the $z$ parameter (number of lines being corrupted) on segmentation accuracy (Dice) and image quality (SSIM). The increasing number of corrupted lines  reduces the performance. Note that the reduction in parameter $z$ refers to a decrease in the number of corrupted lines.}
\label{fig:param_z}
\end{figure}

We also tested the influence of the frame offset for corruption $j$ and the number of lines being corrupted $z$  (see Section \ref{sect:artefactgen}) on the performance. Figure \ref{fig:param_j} illustrates the performance of the algorithm when trained and tested with a fixed offset $j$. Tests were performed for $j=(1,3,5,7,9)$. Similarly we investigated the influence of the number of lines being corrupted to the image quality and segmentation accuracy for $z=(2,4,8,16,32,64)$. These results are shown in Figure \ref{fig:param_z}.

\subsection{Prospective raw k-space data analysis}
\label{sec:ProspectiveResults}

Finally, we investigated the performance of our method in a prospective study, where we acquired fully sampled raw k-space data of mid-ventricular short axis slices with 25 phases from a healthy volunteer under three different circumstances. The first acquisition was a breath-hold image,  cardiac-triggered using the volunteer's ECG signal, which results in a high quality image. Then we acquired 2 different acquisitions with mistriggering artefacts. The mistriggering arose from the use of synthetic ECG signals featuring 60 and 70 beats per minute (bpm) respectively, rather than the real ECG of the volunteer. The results reported below are averaged over these two mistriggered acquisitions. All acquisitions were performed in different breath-holds and a single mid-ventricular slice was acquired.

\begin{figure}[tb]

\begin{minipage}[b]{0.22\linewidth}
  \centering
  \centerline{\includegraphics[trim={1cm 1.5cm 1cm 1.5cm},clip, width=2.0cm]{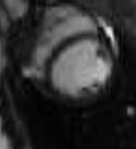}}
%  \vspace{1.5cm}
  \centerline{(a)}\medskip
\end{minipage}
\hfill
\begin{minipage}[b]{0.22\linewidth}
  \centering
  \centerline{\includegraphics[trim={1cm 1.5cm 1cm 1.5cm},clip,width=2.0cm]{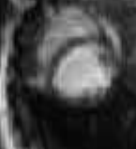}}
%  \vspace{1.5cm}
  \centerline{(b)}\medskip
\end{minipage}
\hfill
\begin{minipage}[b]{0.22\linewidth}
  \centering
  \centerline{\includegraphics[trim={1cm 1.5cm 1cm 1.5cm},clip,width=2.0cm]{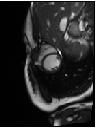}}
%  \vspace{1.5cm}
  \centerline{ (c) } \medskip
\end{minipage}
\hfill
\begin{minipage}[b]{0.22\linewidth}
  \centering
  \centerline{\includegraphics[trim={1cm 1.5cm 1cm 1.5cm},clip,width=2.0cm]{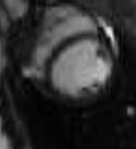}}
%  \vspace{1.5cm}
  \centerline{(d) }\medskip
\end{minipage}
\hfill
\begin{minipage}[b]{0.22\linewidth}
  \centering
  \centerline{\includegraphics[trim={1cm 1.5cm 1cm 1.5cm},clip,width=2.0cm]{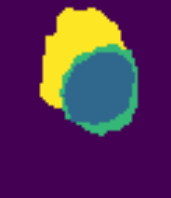}}
%  \vspace{1.5cm}
  \centerline{(e)}\medskip
\end{minipage}
\hfill
\begin{minipage}[b]{0.22\linewidth}
  \centering
  \centerline{\includegraphics[trim={1cm 1.5cm 1cm 1.5cm},clip,width=2.0cm]{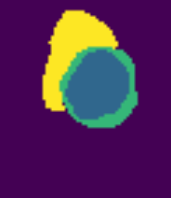}}
  \centerline{(f)} \medskip
\end{minipage}
\hfill
\begin{minipage}[b]{0.22\linewidth}
  \centering
  \centerline{\includegraphics[trim={1cm 1.5cm 1cm 1.5cm},clip,width=2.0cm]{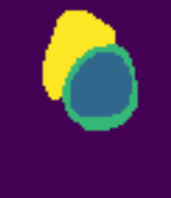}}
%  \vspace{1.5cm}
  \centerline{(g)} \medskip
\end{minipage}
\hfill
\begin{minipage}[b]{0.22\linewidth}
  \centering
  \centerline{\includegraphics[trim={1cm 1.5cm 1cm 1.5cm},clip,width=2.0cm]{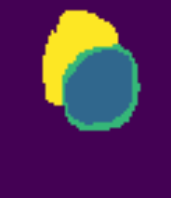}}
%  \vspace{1.5cm}
  \centerline{(h)} \medskip
\end{minipage}
\hfill

\caption{Prospective study results. Artefact are generated by applying mistriggering during acquisition (a), which results in low quality segmentations (e). Raw k-space data is corrected using two U-nets for correction (b) and segmentation (f). Our proposed method produces a high quality image output (c) and a high quality segmentation (g), which is comparable to image acquisition without mistriggering (d) and ground truth segmentation (h).}
\label{fig:raw}
\end{figure}

In Table \ref{table:rawimg}, we compare the image quality results achieved by different algorithms in comparison to our proposed method, in terms of the no-reference image quality metrics SI and SSIM. The baseline refers to the image acquisition with artefacts without applying any correction algorithm. U-net refers to the two consecutive U-net architecture introduced in Section \ref{sec:ArchitectureResults}. The increased image quality is illustrated in Figure \ref{fig:raw}. The high quality image  generated with our framework is similar to the image acquired without mistriggering. This demonstrates that our proposed methodology can be applied on the raw k-space data prospectively and can achieve high image quality.

%that is trained on synthetic data.
In this case, the ground truth segmentation was generated by an expert clinician using the breath-hold image. The baseline refers to the overlap between a manual segmentation of the artefact image and the ground truth on the breath-hold image. The proposed technique achieves high quality segmentation for myocardium, LV and RV blood pool as illustrated in Figure \ref{fig:raw}.

% Prospective Image quality Table
\begin{table} 

\centering
\caption{Image Quality of the algorithm outputs with the good quality acquisition. The trained network achieves high SSIM and sharpness for the data from raw k-space in a prospective study.}
\begin{tabular}{lcc}
\hline
Methods   &  SSIM & SI\\
\hline 
Baseline  & 0.623 & 58.940    \\ 
Cascaded U-nets   & 0.753 & 61.800      \\ 
\textbf{Proposed}  & \textbf{0.793} & \textbf{63.890}   \\ 
\hline
\end{tabular}
\label{table:rawimg}
\end{table}

% Prospective Image Segmentation Table
\begin{table} 

\centering
\caption{Dice segmentation overlap of the algorithm outputs with the ground truth segmentation (manual delineation of the good quality image). The proposed architecture achieves high Dice overlap in prospective setup for  Left Ventricle (LV), Myocardium (Myo) and Right Ventricle (RV) region.}
\begin{tabular}{lccc}
\hline
Methods   &  LV  &  Myo  &  RV \\
\hline 
Baseline  & 0.889 & 0.561  & 0.905  \\ 
Cascaded U-nets   & 0.895 & 0.651  & 0.918    \\ 
\textbf{Proposed}  & \textbf{0.964} & \textbf{0.775} &  \textbf{0.933}   \\ 
\hline
\end{tabular}
\label{table:rawseg}
\end{table}

%%%%%%%%%%%%%%%%%%%%%%%%%%%%%%%%%%%%%%%%%%%%%%%%%%%%%%%%%%%%%%%%%%%%%%%%
% Discussion
%%%%%%%%%%%%%%%%%%%%%%%%%%%%%%%%%%%%%%%%%%%%%%%%%%%%%%%%%%%%%%%%%%%%%%%%
%\newpage
\section{Discussion and conclusion}
\label{sec:discussion_conclusion}

% Work summary
We have presented an extensive study on automatic cardiac motion artefact detection, correction and segmentation in an end-to-end deep learning architecture, which can be used as a global image reconstructor. First, we generated synthetic artefacts from high quality data to train our algorithm and we tested different network architectures to address the tasks of artefact correction and downstream segmentation. Our fundamental contribution in this paper is to address image artefact detection, correction and segmentation jointly, resulting in a network architecture that can output both good quality image reconstructions and segmentations. We have also investigated the use of our algorithm in a prospective study to truly evaluate the clinical potential of our framework.\\

% Results comment
One key observation of our work is the superiority of two consecutive networks for achieving high quality images and image segmentations, compared to 2-channel networks. The end-to-end training of the  serial networks enables them to exploit the interdependencies between the tasks of reconstruction and segmentation, and thereby to produce better quality reconstructed images that are optimised for the downstream task of segmentation. Moreover, we tested various strategies for image artefact correction and proposed a methodology to correct low quality images directly from k-space, which is improved by back-propagating the segmentation loss. It is interesting to observe that a joint network using a combination of detection, correction and segmentation losses improves the performance in terms of both image and segmentation quality. Moreover, the proposed method did not decrease image/segmentation performance on high quality data, which illustrates the success of our method in correctly detecting cases without motion artefacts. Finally, employing a weighting to each loss, the user can tune the network to perform any of the three tasks more accurately, which gives the user the option to choose between high image quality and high segmentation accuracy according to their needs.\\

% Limitations
In future, we would like to include more prospective cases in a real clinical setup to truly evaluate the performance of our algorithm with different MR vendors and field strengths. Moreover, investigation of basal and apical slice quality, which exhibits a slightly different anatomy and challenge, is an important future direction. In addition, 3D processing will require more GPU memory and the architectures should be modified accordingly. In this work, we deliberately used existing network architectures and loss functions to enable us to focus our evaluation on the influence of our joint detection, correction and segmentation framework. One additional avenue of improvement is to investigate novel architectures tailored to the segmentation problem at hand in a multi-task framework.  Finally, our pipeline has so far been tested only on mistriggering artefacts and one future direction is to adjust the network training for respiratory motion artefacts to facilitate the clinical translation of our work under any source of artefact. \\

%  Clinical impact and Future work
The UK Biobank is a controlled study, which uses the general public as a cohort. The real utility of the proposed architecture will be in a clinical setting. Patients who require a CMR scan are likely to have cardiac conditions such as arrythmias, and may have difficulties with breath-holding. Therefore, image artefacts are likely to be more common.
%The proposed algorithm can be utilised to be tuned to get high quality images and segmentations in relation to the clinical necessity and application. Moreover, our technique offers a global image reconstructor and consistency, which is highly desirable for CMR.
With the successful translation of our architecture in clinical setups, high diagnostic image and segmentation quality can be offered consistently. Furthermore, MR images of other organs can suffer from artefacts and require accurate segmentations, and we believe that a similar paradigm could be applied for those.\\

% Conclusion 
In conclusion, this work represents an important contribution to CMR image reconstruction. Our novel idea of detecting and correcting artefacts from k-space  for high segmentation accuracy has been shown to improve both reconstructed image quality and segmentation quality.  In the current environment of the increasing use of imaging in clinical practice, as well as the emergence of large population data cohorts which include imaging, our proposed global image reconstructor and segmenter can ensure high quality outputs independent of such motion artefacts.

% ********* / PAPER CORE *********

\ifCLASSOPTIONcaptionsoff
 \newpage
\fi

% ********* REFERENCES SECTION *********
%
\bibliographystyle{IEEEtran}
\bibliography{TMI19_1}

\end{document}